\documentclass[letterpaper, 10 pt, conference]{ieeeconf}  % Comment this line out if you need a4paper

\IEEEoverridecommandlockouts                              % This command is only needed if 
\usepackage[latin1]{inputenc} 
\usepackage{fancyhdr}
\usepackage[hang]{caption2}
\usepackage[english]{babel}

\usepackage{epsfig}
\usepackage{graphicx}
\usepackage[normalsize,it,hang]{subfigure}
\usepackage{amsfonts,amsmath,bm,hhline}

\usepackage{color}
\newtheorem{remark}{Remark}

\title{\LARGE \bf
A simple but energy-efficient HVAC control synthesis for data centers}

\author{Michel Fliess$^{1,3}$, C\'{e}dric Join$^{2,3}$, Maria Bekcheva$^{4}$, Alireza Moradi$^{4}$, Hugues Mounier$^{5}$
\thanks{$^{1}$LIX (CNRS, UMR 7161), \'Ecole polytechnique, 91128 Palaiseau, France, Michel.Fliess@polytechnique.edu}
\thanks{$^{2}$CRAN (CNRS, UMR 7039), Universit\'{e} de Lorraine, BP 239, 54506 Vand{\oe}uvre-l\`{e}s-Nancy, France, cedric.join@univ-lorraine.fr}
\thanks{$^{3}$AL.I.E.N. (ALg\`{e}bre pour Identification \& Estimation Num\'{e}riques), 7 rue Maurice Barr\`{e}s, 54330 V\'{e}zelise, France, \newline \{michel.fliess, cedric.join\}@alien-sas.com}
\thanks{$^{4}$Inagral, 128 rue  de la Bo\'{e}tie, 75008 Paris, France, \newline \{maria, alireza@inagral.com\}} 
\thanks{$^{5}$Laboratoire des Signaux et Syst\`emes (L2S),  Universit\'e Paris-Sud-CNRS-CentraleSup\'elec, Universit\'e Paris-Saclay, 
91192 Gif-sur-Yvette, France, hugues.mounier@l2s.centralesupelec.fr}
}

\begin{document}

\maketitle
\thispagestyle{empty}
\pagestyle{empty}

%%%%%%%%%%%%%%%%%%%%%%%%%%%%%%%%%%%%%%%%%%%%%%%%%%%%%%%%%%%%%%%%%%%%%%%%%%%%%%%%
\begin{abstract}
The air conditioning management of data centers, a key question with respect to energy saving, is here tackled via the recent model-free control synthesis. Mathematical modeling becomes useless in this approach. The tuning of the corresponding intelligent 
proportional controller is straightforward. Computer simulations show excellent tracking performances in various realistic situations, like CPU load or temperature changes.

\end{abstract}

%%%%%%%%%%%%%%%%%%%%%%%%%%%%%%%%%%%%%%%%%%%%%%%%%%%%%%%%%%%%%%%%%%%%%%%%%%%%%%%%
\section{Introduction}
Two exciting advances in cloud computing \cite{berkeley}, a fast growing industry in information technology, have been recently derived by the authors:
\begin{itemize}
\item improving resource elasticity \cite{iste} thanks to model-free control in the sense of \cite{csm},
\item workload forecasting \cite{codit} via time series analysis as in \cite{solar}.
\end{itemize}
Data centers, which are fundamental in this context,  consume a huge amount of electrical energy \cite{bel,jones,wu}. Almost half of it is devoted to their cooling. The aim of this communication is to show  that model-free control might provide also a most efficient control tool with respect to air conditioning. 
\begin{remark}
\begin{enumerate}
\item \emph{HVAC}, \textit{i.e.}, \emph{heating, ventilation, and air conditioning},  which is defined by Wikipedia as ``the technology of indoor and vehicular environmental comfort'' (see, \textit{e.g.}, \cite{ka}), plays therefore a key r\^{o}le (see,  \textit{e.g.}, \cite{capo,joshi}). The corresponding numbers of publications and patents are increasing rapidly. 
\item From an applied control engineering perspective, on/off  and PID controllers seem to be widely used (see, \textit{e.g.}, \cite{benoit,durand1,durand2,lee,qu,terra}, and the references therein). To a large extent this situation is explained by their conceptual simplicity. Nevertheless their tuning, which is too often a quagmire, might lead to poor performances. 
\item Most of the model-based approaches rest on various optimization techniques (see, \textit{e.g.}, \cite{beghi,conficoni,jiang,pakb,rong,wang}). Let us add that predictive control  (see, \textit{e.g.}, \cite{cupelli,fang,lazic,parolini,paul,yao,zhou}) is essential in that respect. Deriving sound mathematical modeling  necessitates complex parameter identification and/or machine learning procedures in order to get convincing results (see, \textit{e.g.}, \cite{fu}). 
\end{enumerate}
\end{remark} 
\begin{remark}
Besides excellent existing results on the HVAC of greenhouses \cite{toulon} and buildings \cite{roumanie,bara2,micha,telsang}, model-free control has already given birth to many successful concrete applications (see the references in \cite{csm} and \cite{bara1,logistique}) including some patents.
\end{remark}

Our paper is organized as follows. Model-free control is summarized in Section \ref{mfc}. A simplified mathematical modeling via ordinary differential equations is sketched in Section \ref{modeling} for the purpose of computer simulations. 
The performances of our control synthesis are presented and discussed in Section  \ref{simu}. Section \ref{con} is devoted to some concluding remarks.

\section{What is model-free control?\protect\footnote{See \cite{csm} for more details.}}\label{mfc}
\subsection{Generalities}
Replace the unknown or poorly known SISO system by \emph{ultra-local} model 
\begin{equation}
{\dot{y} = F + \alpha u} \label{1}
\end{equation}
where
\begin{itemize}
	\item $u$ and $y$ are the input (control) and output variables,
	\item the derivation order of $y$ is $1$, like in most concrete situations,
	\item $\alpha \in \mathbb{R}$ is chosen by the practitioner such that 
	$\alpha u$ and $\dot{y}$ are of the same magnitude.
\end{itemize}
%\begin{remark}
%	Note that the fact of using an ultra-local model does not compromise the essence of the control approach, where the meaning of model free, refers to making dispensable the task of doing modeling and deriving the governing dynamical equation. 
%\end{remark}
%\vspace{2mm}
The following explanations on $F$ might be useful: 
\begin{itemize}
\item $F$ subsumes the knowledge of any model uncertainties and disturbances,
	\item $F$ is estimated via the measures of $u$ and $y$.
	\end{itemize}

\subsection{Intelligent controllers}
The loop is closed by an \emph{intelligent proportional controller}, or \emph{iP},
\begin{equation}\label{ip}
u = - \frac{\hat{F} - \dot{y}^\ast + K_P e}{\alpha}
\end{equation}
where
\begin{itemize}
	\item $y^\star$ is the reference trajectory,
	\item $e = y - y^\star$ is the tracking error,
	\item $K_P$ is the usual tuning gain.
\end{itemize}
Combining equations \eqref{1} and \eqref{ip} yields:
$$
	\dot{e} + K_P e = 0
$$
where $F$ does not appear anymore. Local exponential stability is ensured if $Kp>0$: 
\begin{itemize}
\item The gain $K_P$ is thus easily tuned. 
\item Robustness with respect to different types of disturbances and model uncertainties is achieved.
\end{itemize}
\begin{remark}
See \cite{csm} for a discussion about the equivalence between the iP \eqref{ip} and proportional-integral controllers (PIs).
\end{remark}
\subsection{Real-time estimation of $F$}\label{F}
The term   $F$ in Equation \eqref{1} is estimated in real-time  according to recent algebraic identification techniques \cite{sira}. 
It may be assumed to be ``well'' approximated by a piecewise constant function $\hat{F}$ (see, \textit{e.g.}, \cite{bourbaki}). Rewrite then Equation \eqref{1}  in the operational domain (see, \textit{e.g.}, \cite{yosida}): 

\begin{align}
	sY = \frac{\Phi}{s}+\alpha U +y(0)
\end{align}

\noindent where $\Phi$ is a constant. We get rid of the initial condition $y(0)$ by multiplying both sides on the left by $\frac{d}{ds}$:
\begin{align}
Y + s\frac{dY}{ds}=-\frac{\Phi}{s^2}+\alpha \frac{dU}{ds}
\end{align}
Noise attenuation is achieved by multiplying both sides on the left by $s^{-2}$. It yields in the time domain the real-time estimate, thanks to the equivalence between $\frac{d}{ds}$ and the multiplication by $-t$,
\begin{equation}\label{integral1}
{\small \hat{F}(t)  =-\frac{6}{\tau^3}\int_{t-\tau}^t \left\lbrack (\tau -2\sigma)y(\sigma)+\alpha\sigma(\tau -\sigma)u(\sigma) \right\rbrack d\sigma }
\end{equation}
where $\tau > 0$ might be quite ``small.''

\section{A simple mathematical model for \\ computer simulations}\label{modeling}
\begin{figure*}[!ht]
\centering%
\subfigure[\footnotesize Air flow]
{\epsfig{figure=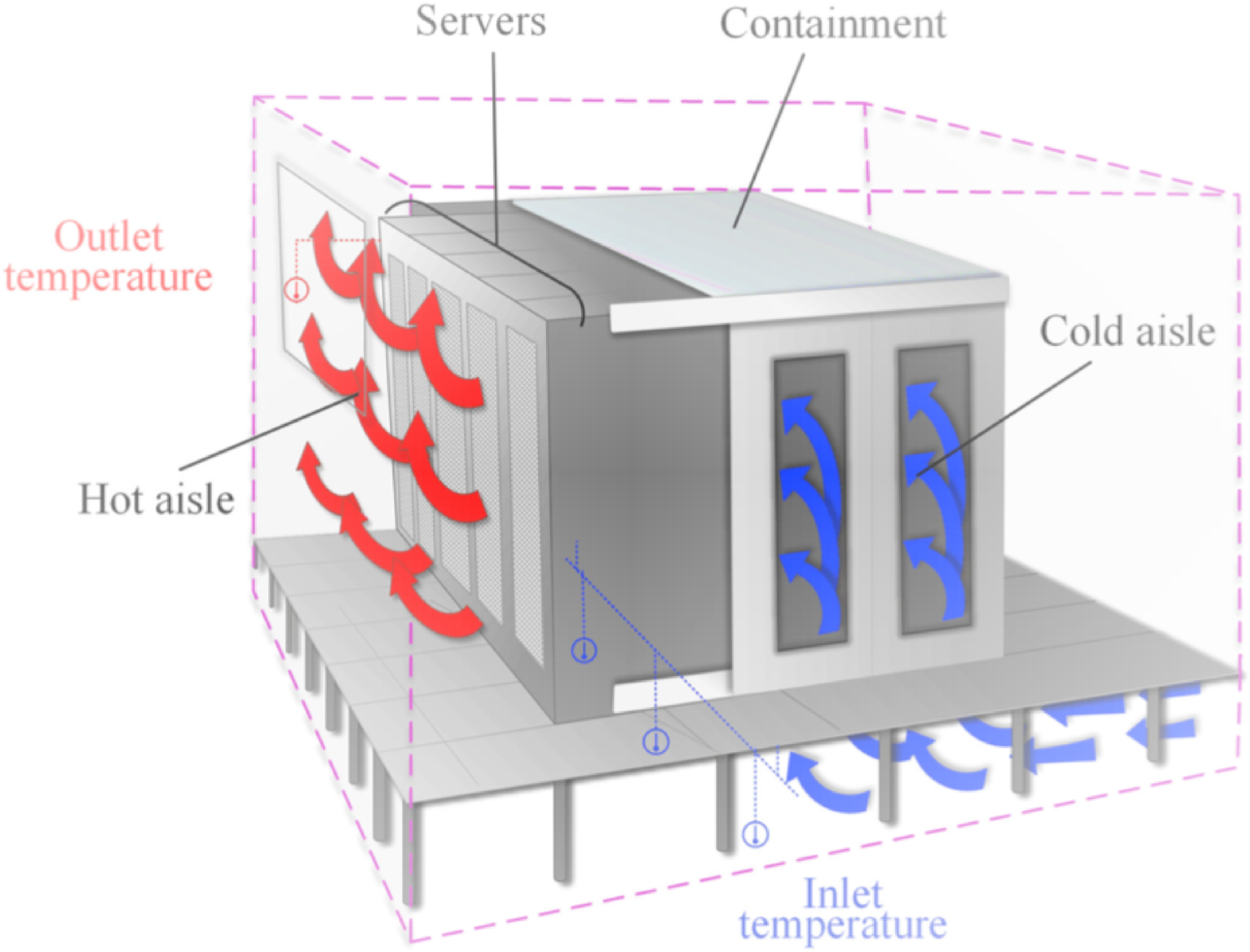,width=0.28\textwidth}}%\hspace{.5cm}
%
%\\
\subfigure[\footnotesize Pictures]
{\epsfig{figure=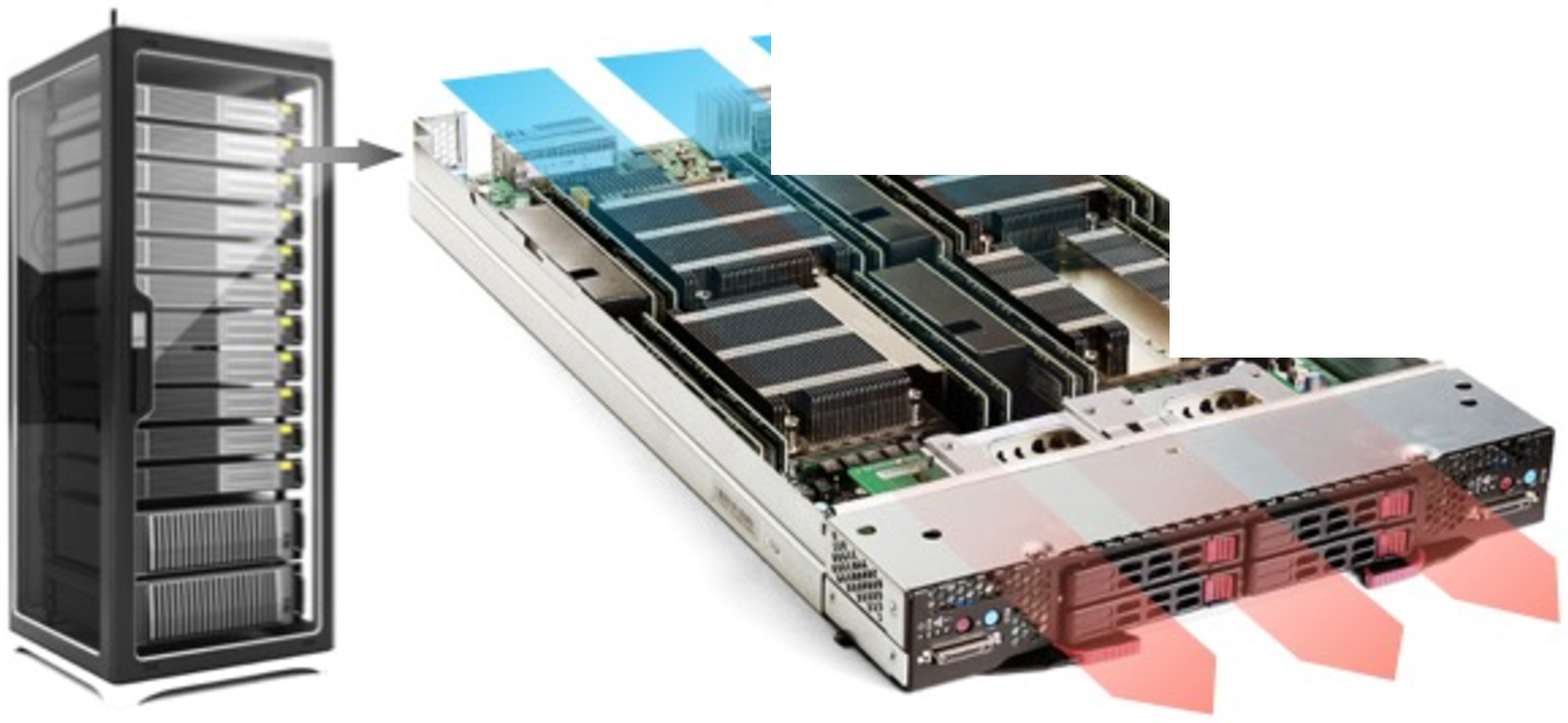,width=0.37\textwidth}}
\caption{A simplified data center}
\label{Bed}
\end{figure*}

\begin{figure*}[!ht]
\centering%
%\subfigure[\footnotesize Commande]
\includegraphics[width=0.68\textwidth]{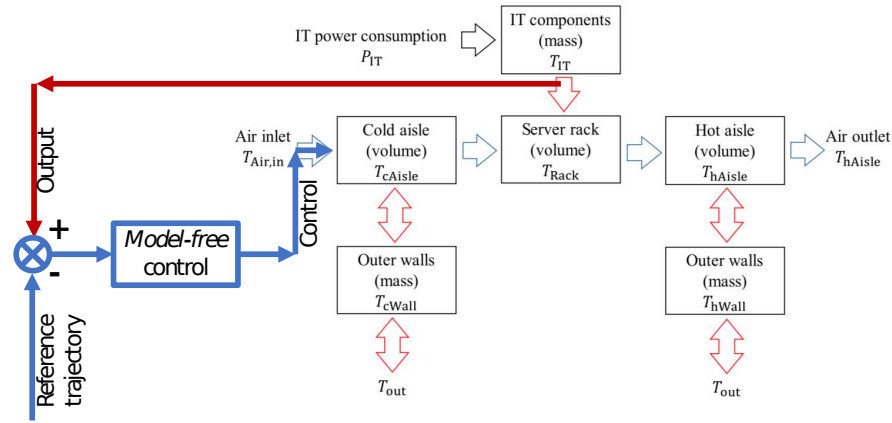}
%\vspace{-0.1cm}
\caption{Control scheme}\label{Flux}
\end{figure*}

Our model, which is essential for computer simulations, is to a great extent borrowed from \cite{cupelli}. 
Figures \ref{Bed}-(a) and \ref{Bed}-(b) represent respectively the server air flow circulation and the simplified data center. Figure \ref{Flux} is sketching the controller and permits to define various important variables. Basic thermodynamic laws lead to the differential equations

%\begin{figure*}[!ht]
%\centering%
%%\subfigure[\footnotesize Commande]
%\includegraphics[width=.4\textwidth]{TestBed.pdf}
%\caption{Sketch of a simple data center}\label{Bed}
%\end{figure*}

%\begin{figure*}[!ht]
%\centering%
%\subfigure[\footnotesize Air flow]
%{\epsfig{figure=TestBed.pdf,width=0.32\textwidth}}\hspace{.5cm}

%\\
%\subfigure[\footnotesize Pictures]
%{\epsfig{figure=ServerAirFlow.pdf,width=0.32\textwidth}}
%\caption{A simplified data center}
%\label{Bed}
%\end{figure*}

%ServerAirFlow.pdf

\begin{equation}
\label{syst}
\begin{cases}
\dot T_{\text{IT}}=\alpha_{\rm 11}P_{\text{IT}}-\alpha_{\rm 12}(T_{\rm IT}-T_{\text{Rack}})\\
\dot T_{\rm Rack}=\alpha_{\rm 21}(T_{\rm cAisle}-T_{\rm Rack})+\alpha_{\rm 22}(T_{IT}-T_{\rm Rack})\\
\dot T_{\rm cAisle}=\alpha_{\rm 31}(T_{\rm Air,in}-T_{\rm cAisle})+\alpha_{\rm 32}(T_{\rm cAisle}-T_{\rm cWall})\\ 
\dot T_{\rm cWall}=\alpha_{\rm 41}(T_{\rm out}-T_{\rm cWall})+\alpha_{\rm 42}(T_{\rm cWall}-T_{\rm cAisle})\\ 
\dot T_{\rm hAisle}=\alpha_{\rm 51}(T_{\rm Rack}-T_{\rm hAisle})+\alpha_{\rm 52}(T_{\rm hAisle}-T_{\rm hWall})\\
\dot T_{\rm hWall}=\alpha_{\rm 61}(T_{\rm out}-T_{\rm hWall})+\alpha_{\rm 62}(T_{\rm hWall}-T_{\rm hAisle})\\ 
\end{cases}
\end{equation}
where
\begin{itemize}
\item $P_{\rm IT}$ is the input power which corresponds to the CPU load, 
\item $T_{\rm Air,in}$ (resp. $T_{\rm IT}$) is the control (resp. output) variable,
\item $\alpha_{\rm ij}$ are suitable parameters.
\end{itemize}
From a classic control-theoretic viewpoint,
\begin{itemize}
\item Equations \eqref{syst} yields a system ($\Sigma$) with a single input $u = T_{\rm Air,in}$ and a single output $y = T_{\rm IT}$ (see also Figure \ref{Flux}),
\item $P_{\rm IT}$ may be viewed as an external disturbance.
\end{itemize}

\section{Some computer simulations}\label{simu}
\subsection{Basic facts}\label{basic}
The following values of the parameters are inspired by \cite{cupelli}: $\alpha_{\rm 11}=2.7248$, $\alpha_{\rm 12}=-32.6975$, $\alpha_{\rm 21}=4.2997.10^3$, $\alpha_{\rm 22}=2.9632.10^4$, $\alpha_{\rm 31}=537.4670$, $\alpha_{\rm 32}=131.6406$, 
$\alpha_{\rm 41}=514.2857$, $\alpha_{\rm 42}=153.5354$, $\alpha_{\rm 51}=335.9169$, $\alpha_{\rm 52}=7.7166$, $\alpha_{\rm 61}=12$, $\alpha_{\rm 62}=9.6000$.  Following again \cite{cupelli}, the output $y$ of System ($\Sigma$) is assumed to track the setpoint $20.9°$ (degree Celsius). In Formulae \eqref{1}-\eqref{ip}, set  $\alpha=10$, $K_p= 1$. The sampling period is $1$ min.

\begin{remark}
Note that forecasting results via time series were used in \cite{cupelli}. They become pointless here.
\end{remark}

\subsection{Four preliminary scenarios}
\subsubsection{Sudden CPU load change}
\begin{figure*}[!ht]
\centering%
\subfigure[\footnotesize Power consumption: CPU load $P_{IT}$  (kW) ]
{\epsfig{figure=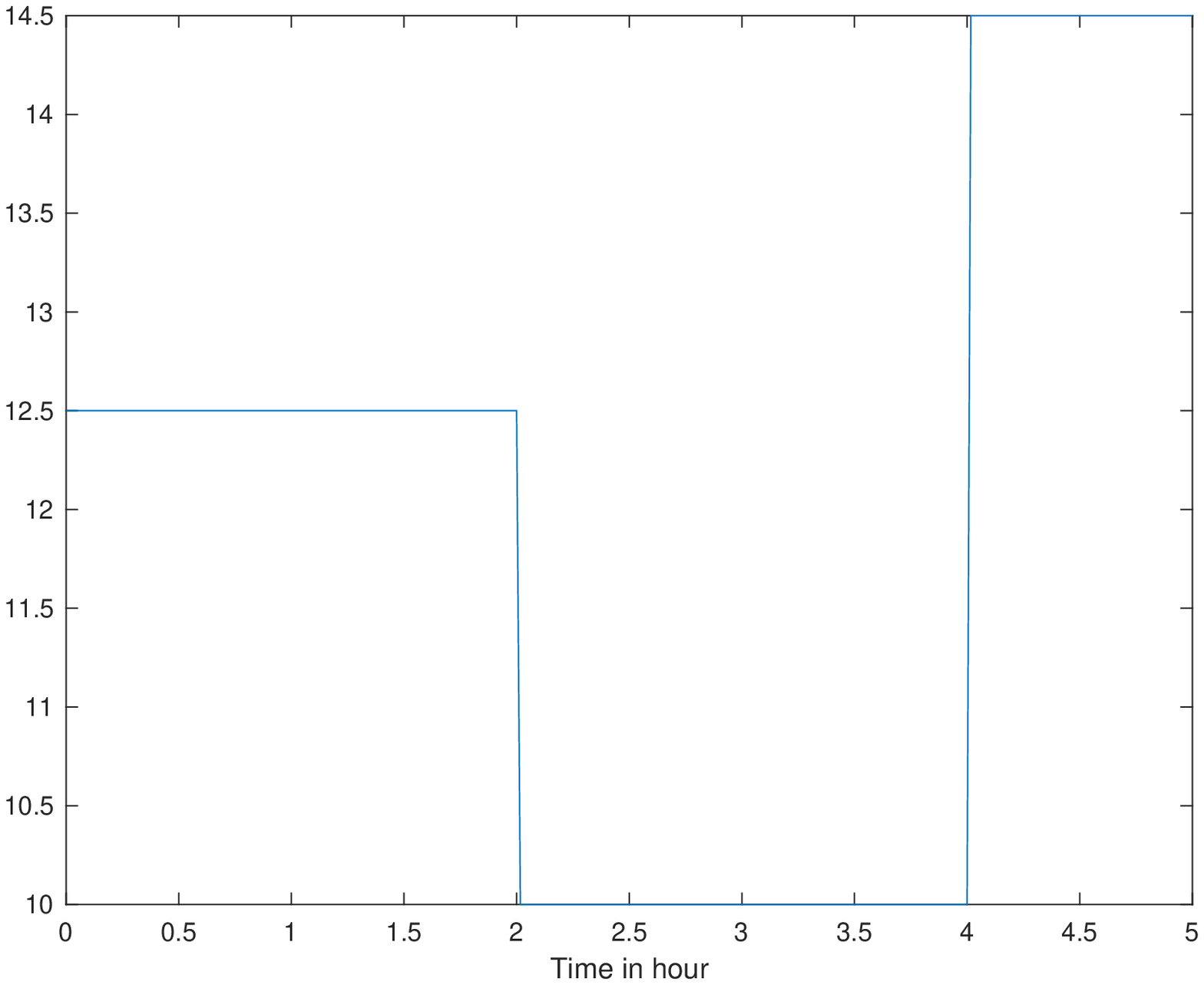,width=0.30\textwidth}}%\hspace{.5cm}
%
%\\
\subfigure[\footnotesize Data center emperature $T_{out}$ (degree Celsius)]
{\epsfig{figure=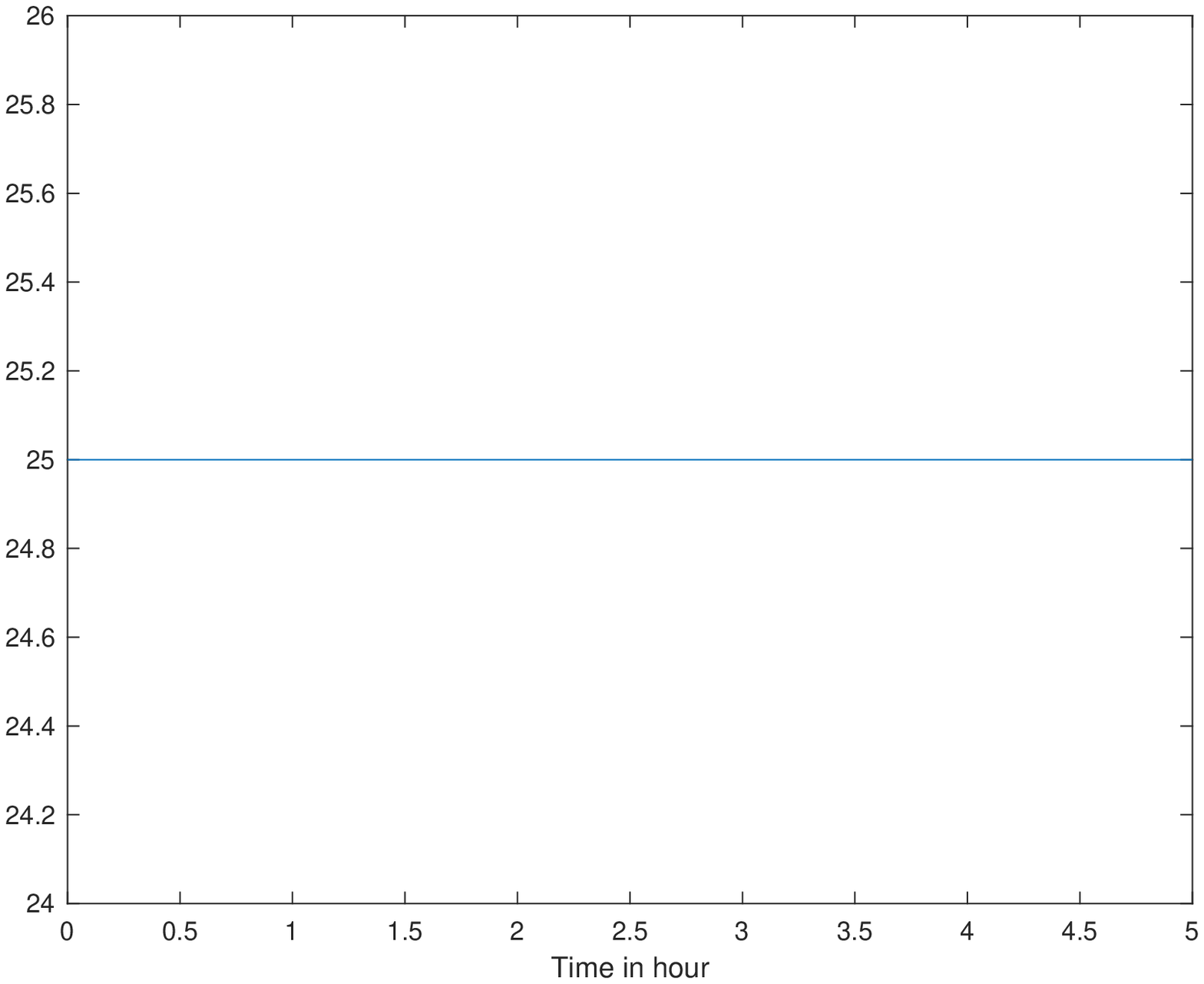,width=0.30\textwidth}}%\hspace{.5cm}
\\
\subfigure[\footnotesize Control variable $T_{Air,in}$  (degree Celsius) ]
{\epsfig{figure=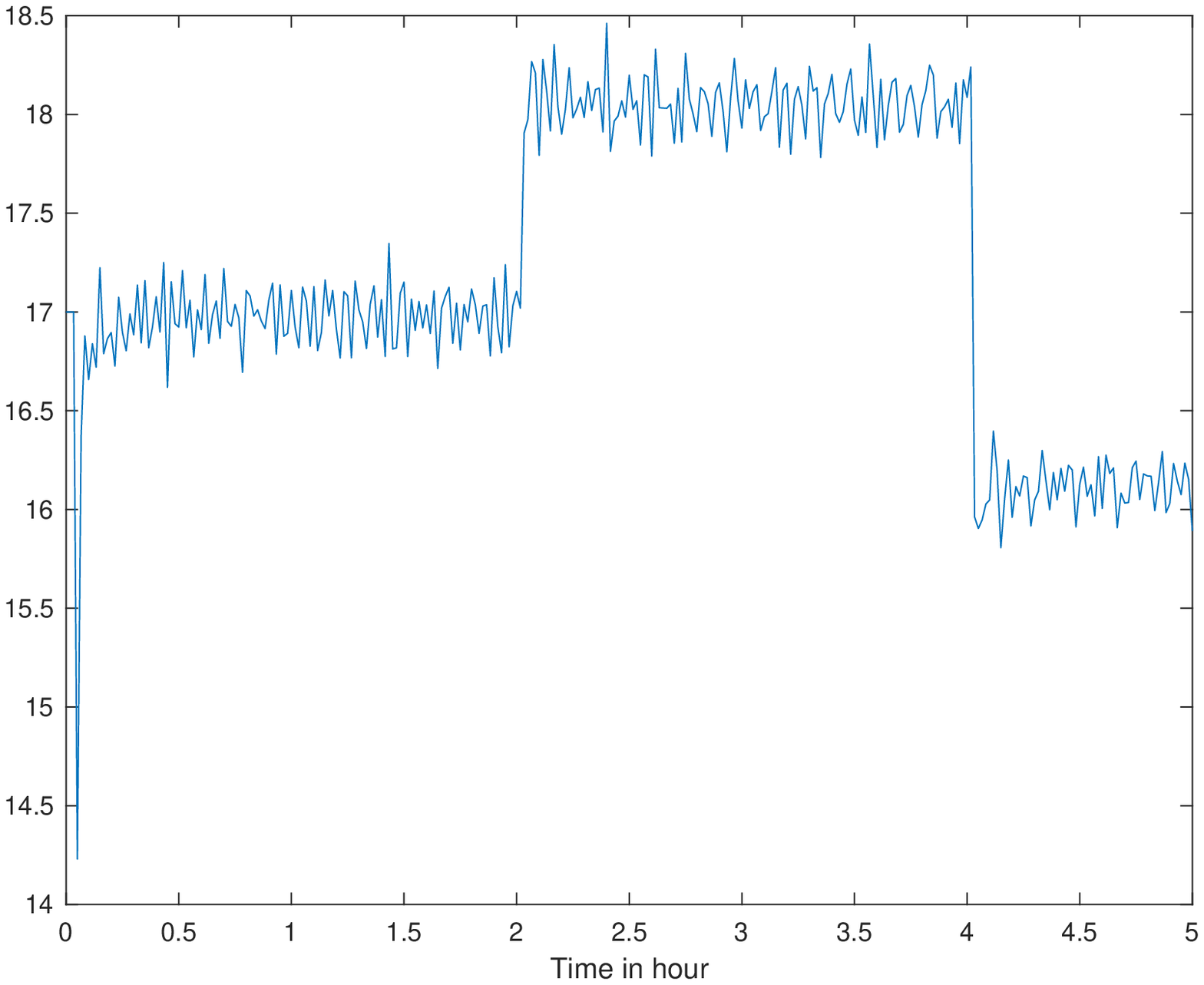,width=0.30\textwidth}}%\hspace{.5cm}
\subfigure[\footnotesize Output variable $T_{IT}$  (degree Celsius) ]
{\epsfig{figure=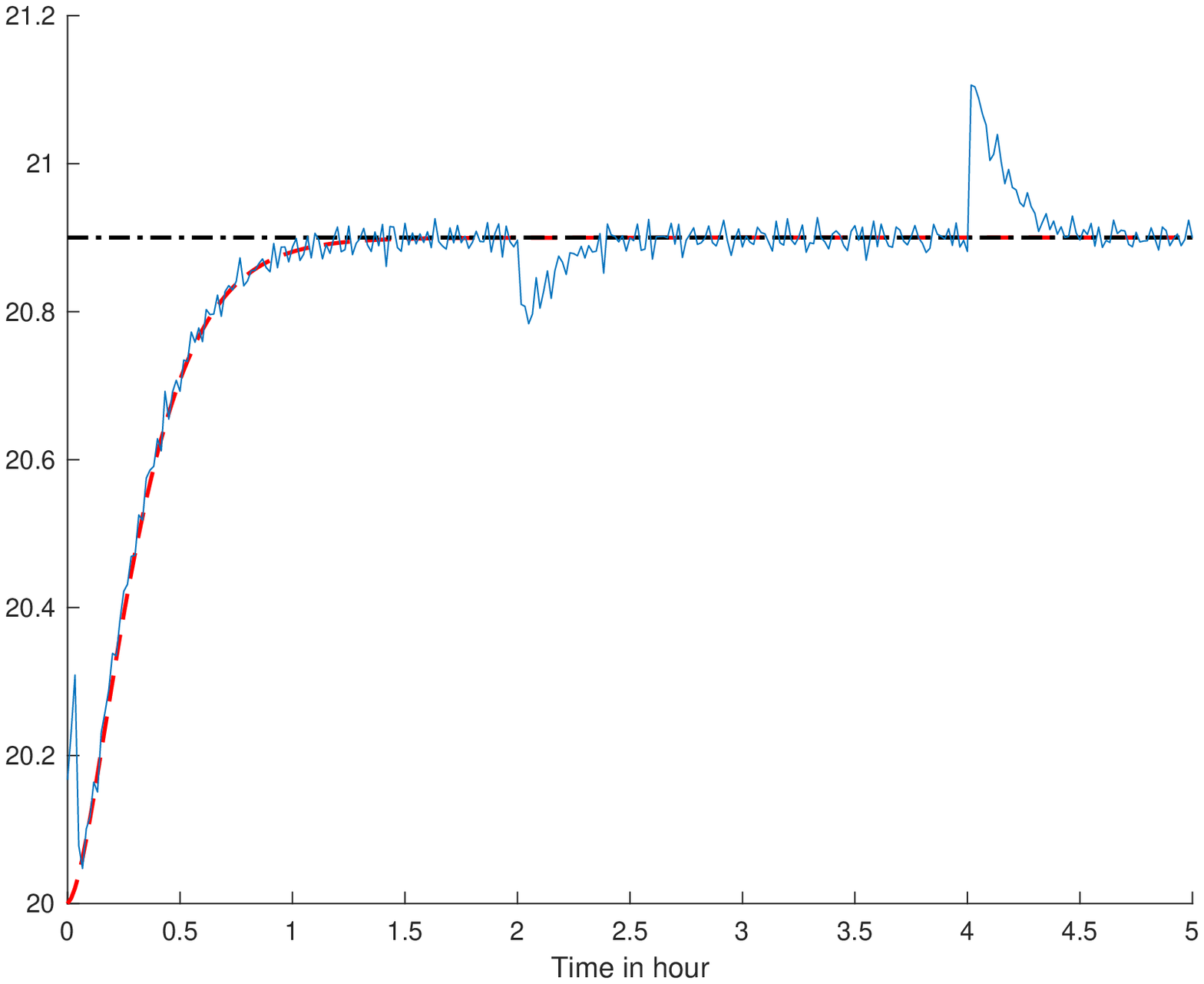,width=0.30\textwidth}}%\hspace{.5cm}
\caption{Sudden CPU load change}\label{S2}
\end{figure*}

Figure \ref{S2}-(a) exhibits a sudden change of the CPU load $P_{\rm IT}$. Figure \ref{S2}-(d) shows that the setpoint is well tracked.

%%%%%%%%%%%%%%%%
% Ajout de l'oubli %
%%%%%%%%%%%%%%%%
\subsubsection{A more realistic CPU load change}
\begin{figure*}[!ht]
\centering%
\subfigure[\footnotesize Power consumption: CPU load $P_{IT}$  (kW) ]
{\epsfig{figure=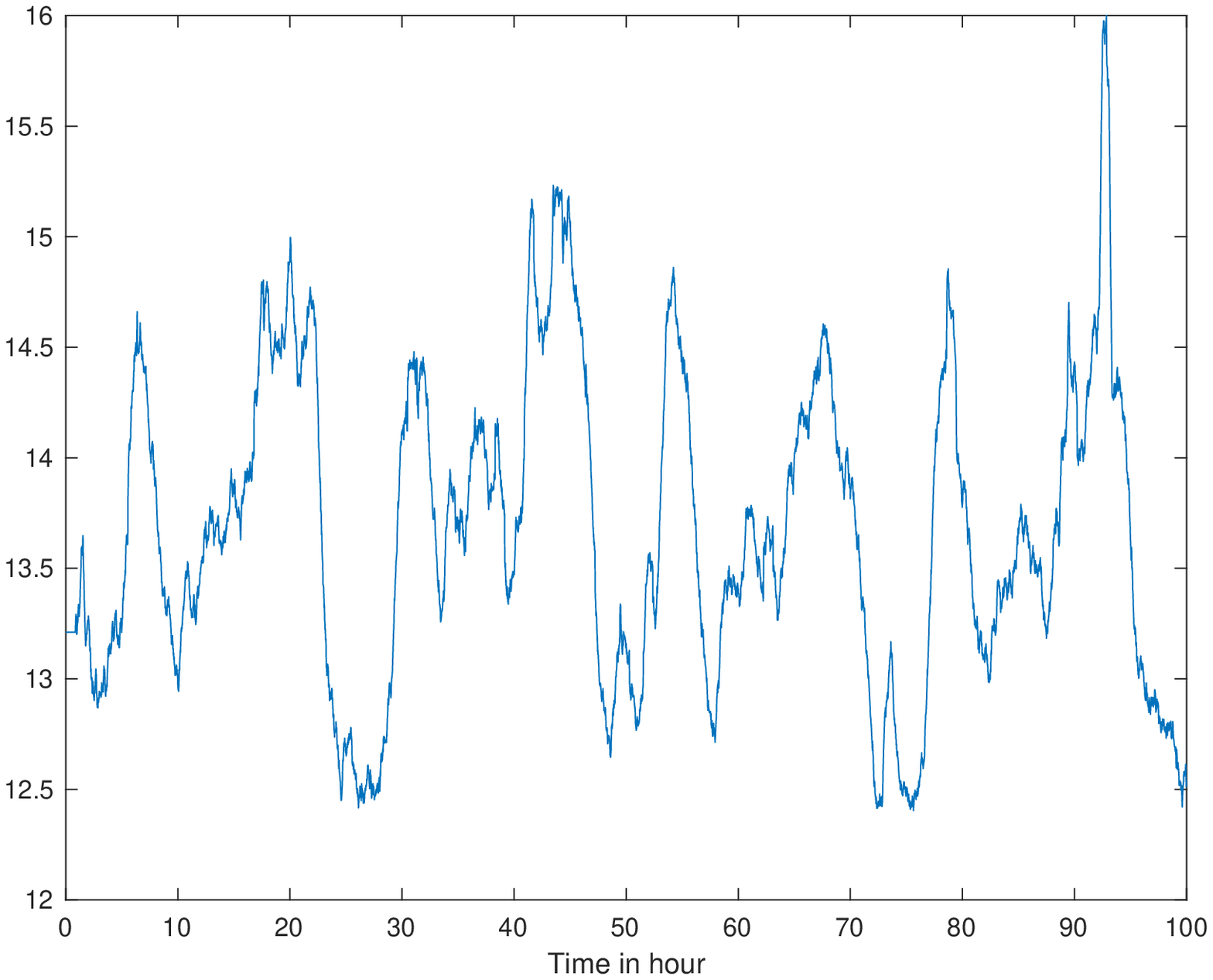,width=0.30\textwidth}}%\hspace{.5cm}
%
%\\
\subfigure[\footnotesize Data center temperature $T_{out}$ (degree Celsius)]
{\epsfig{figure=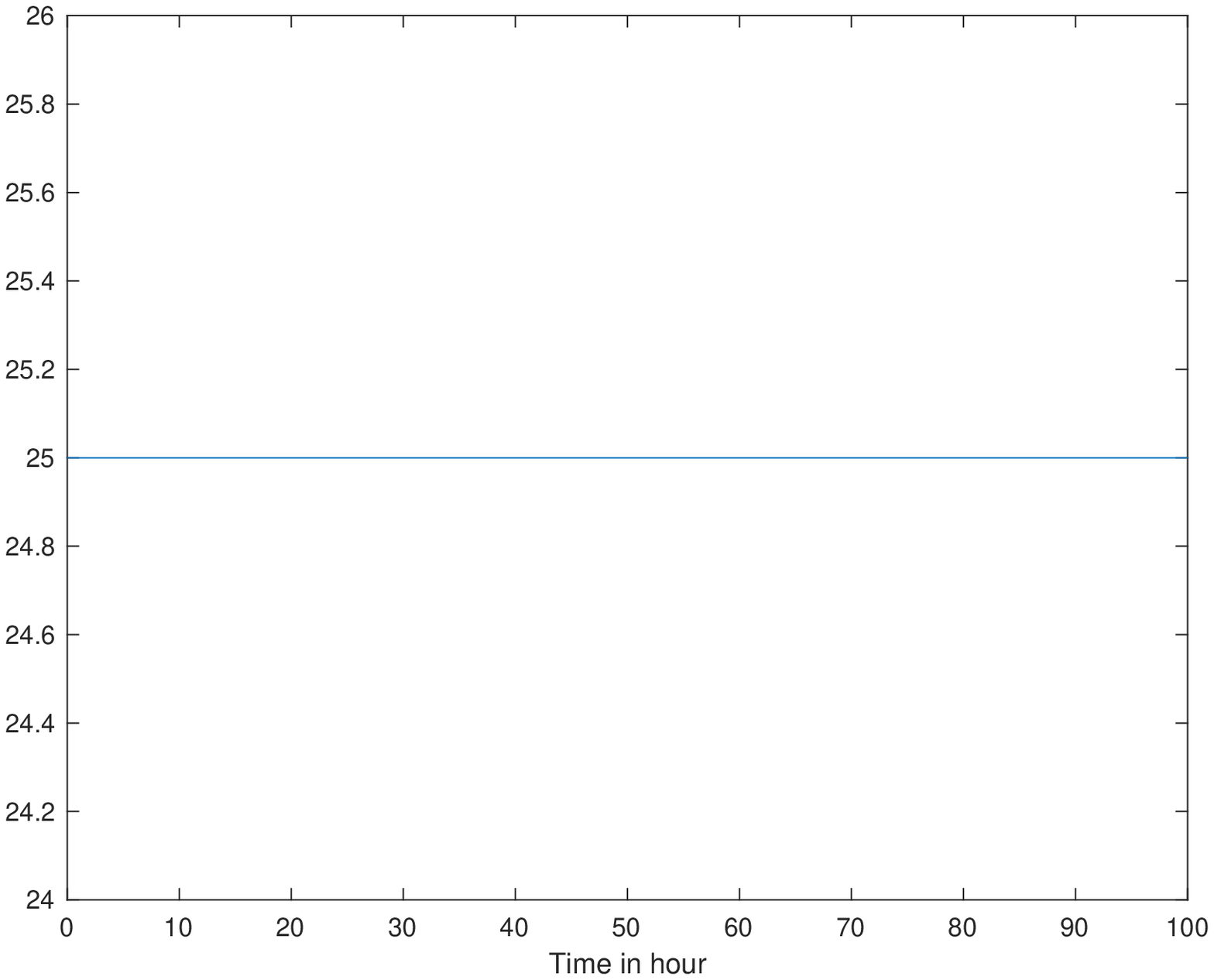,width=0.30\textwidth}}%\hspace{.5cm}
\\
\subfigure[\footnotesize Control variable $T_{Air,in}$  (degree Celsius) ]
{\epsfig{figure=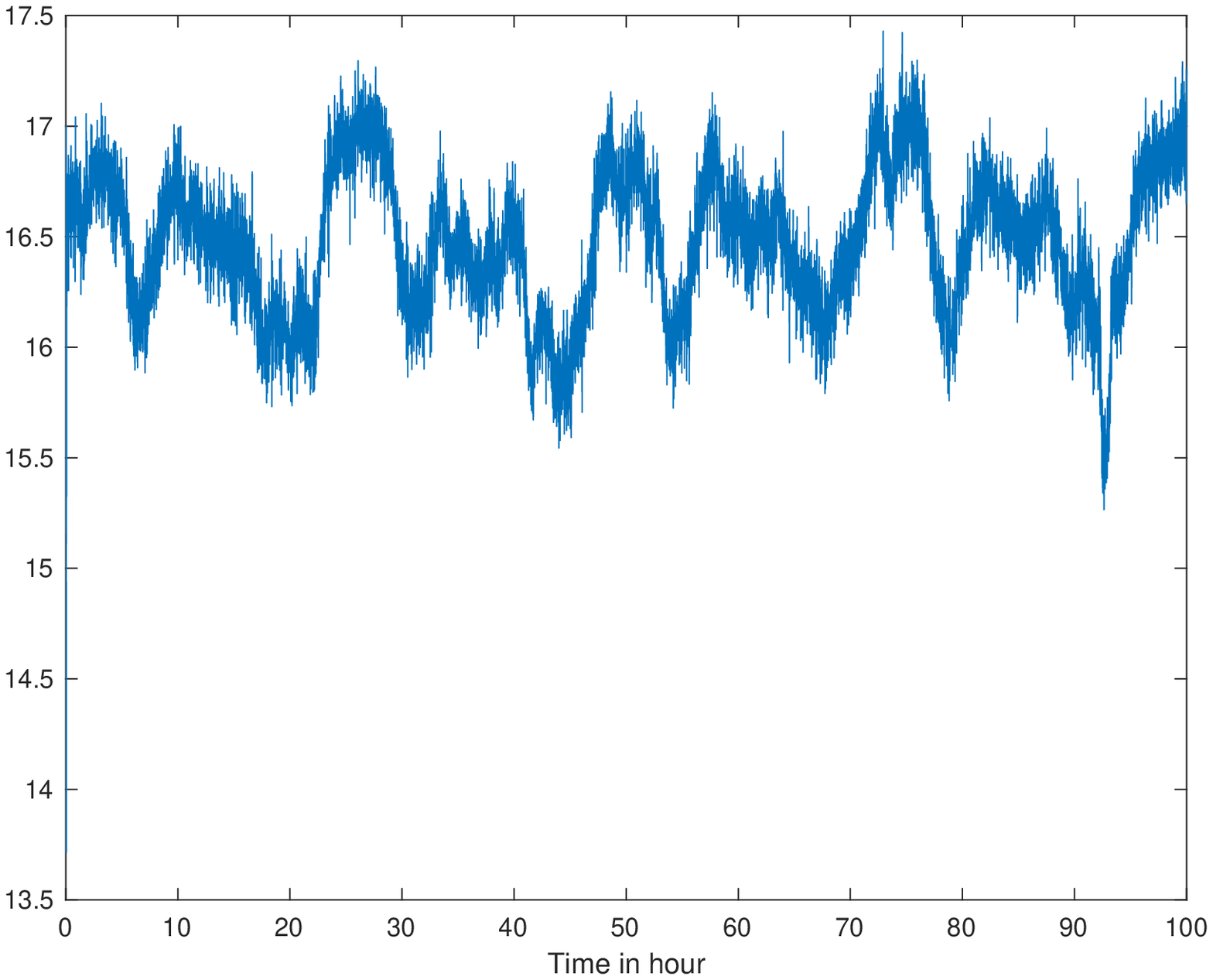,width=0.30\textwidth}}%\hspace{.5cm}
\subfigure[\footnotesize Output variable $T_{IT}$  (degree Celsius) ]
{\epsfig{figure=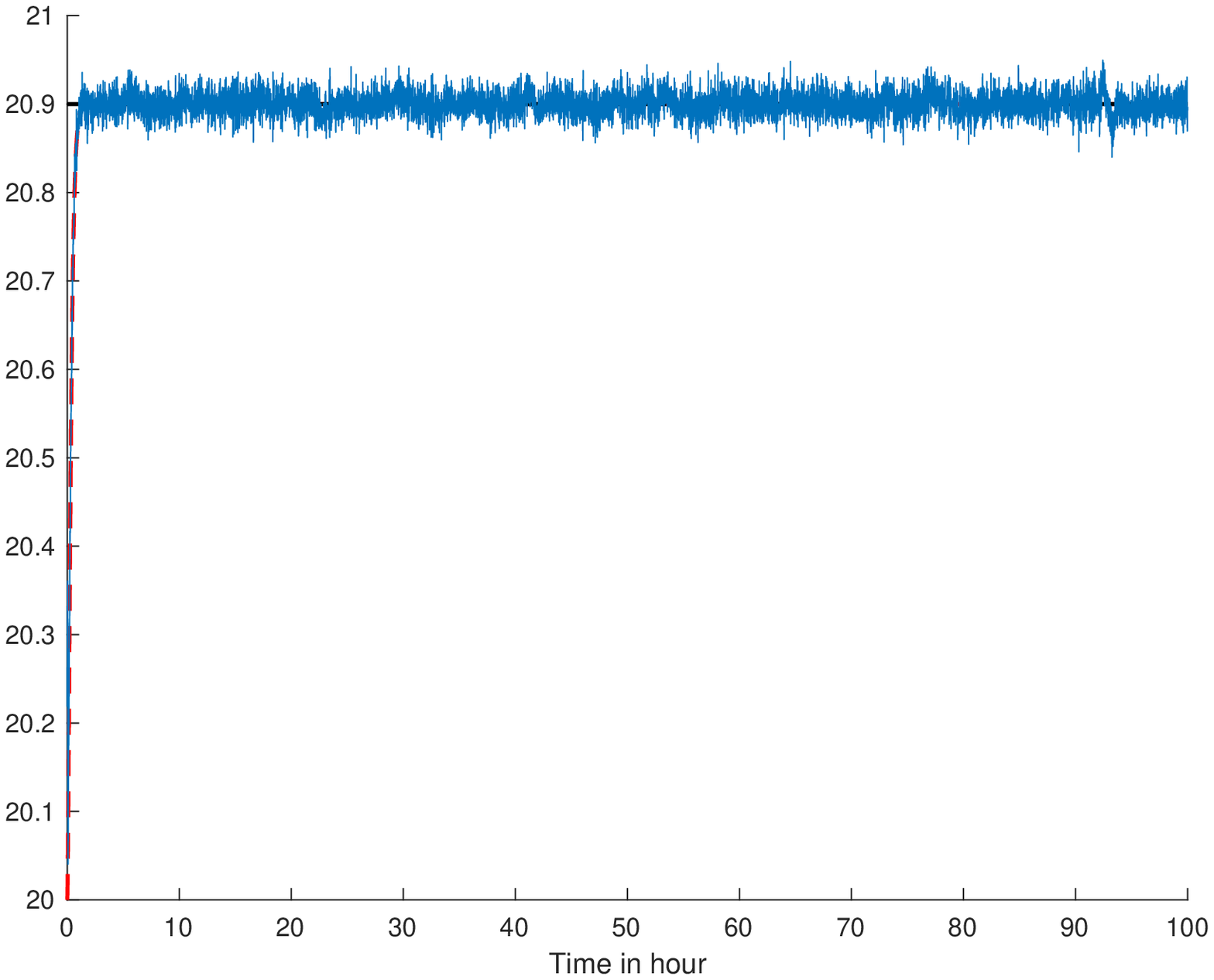,width=0.30\textwidth}}%\hspace{.5cm}
\caption{Realistic CPU load change}\label{S3}
\end{figure*}

It is given by lnagral, \textit{i.e.}, the Company to which two authors, M. Bekcheva and A. Moradi, belong, and is depicted in Figure \ref{S3}-(a).
Figures \ref{S3}-(d) confirms a great tracking. 

\subsubsection{Sudden temperature change}
\begin{figure*}[!ht]
\centering%
\subfigure[\footnotesize Power consumption: CPU load $P_{IT}$  (kW) ]
{\epsfig{figure=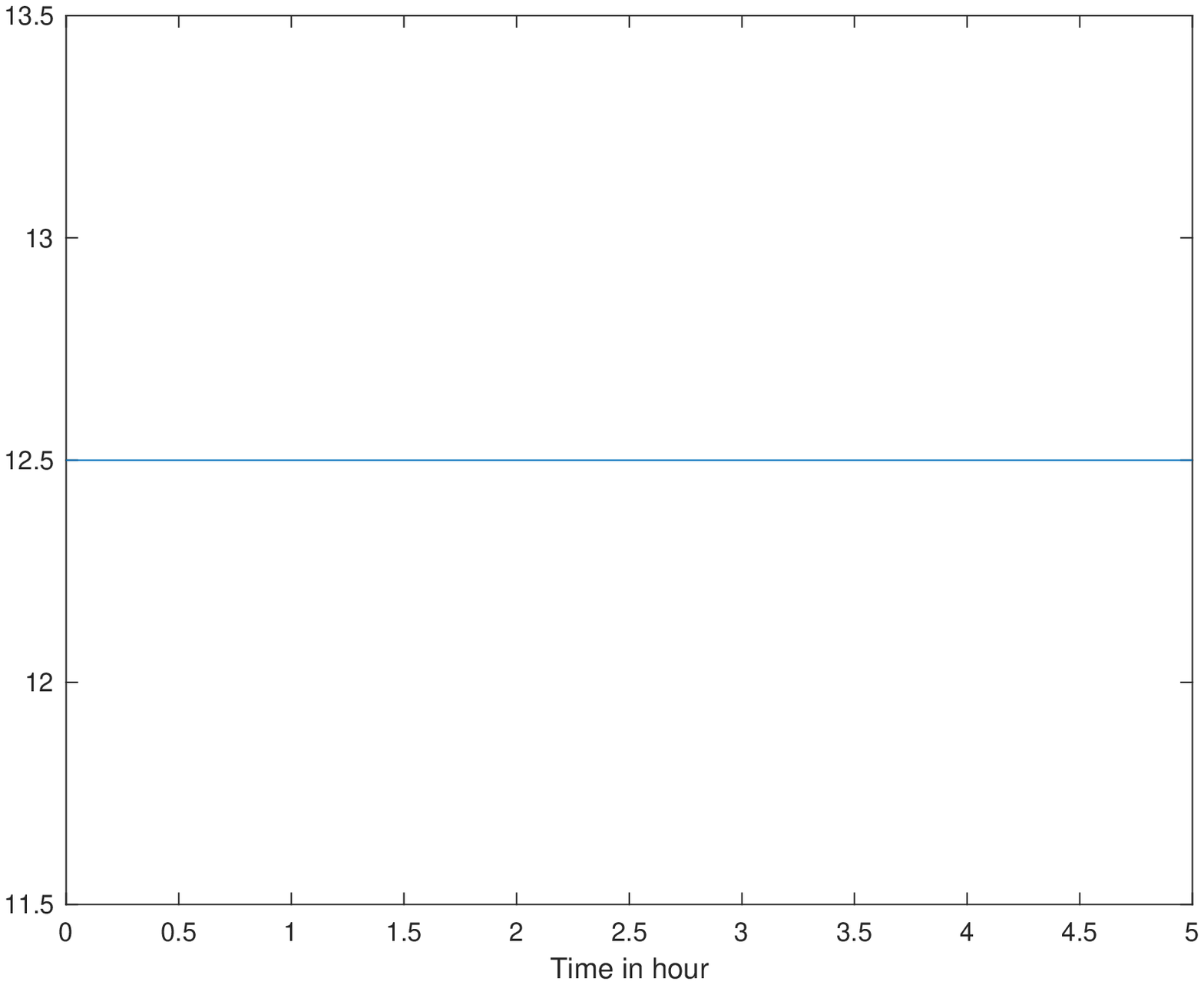,width=0.30\textwidth}}%\hspace{.5cm}
%
%\\
\subfigure[\footnotesize Data center temperature  $T_{out}$ (degree Celsius)  ]
{\epsfig{figure=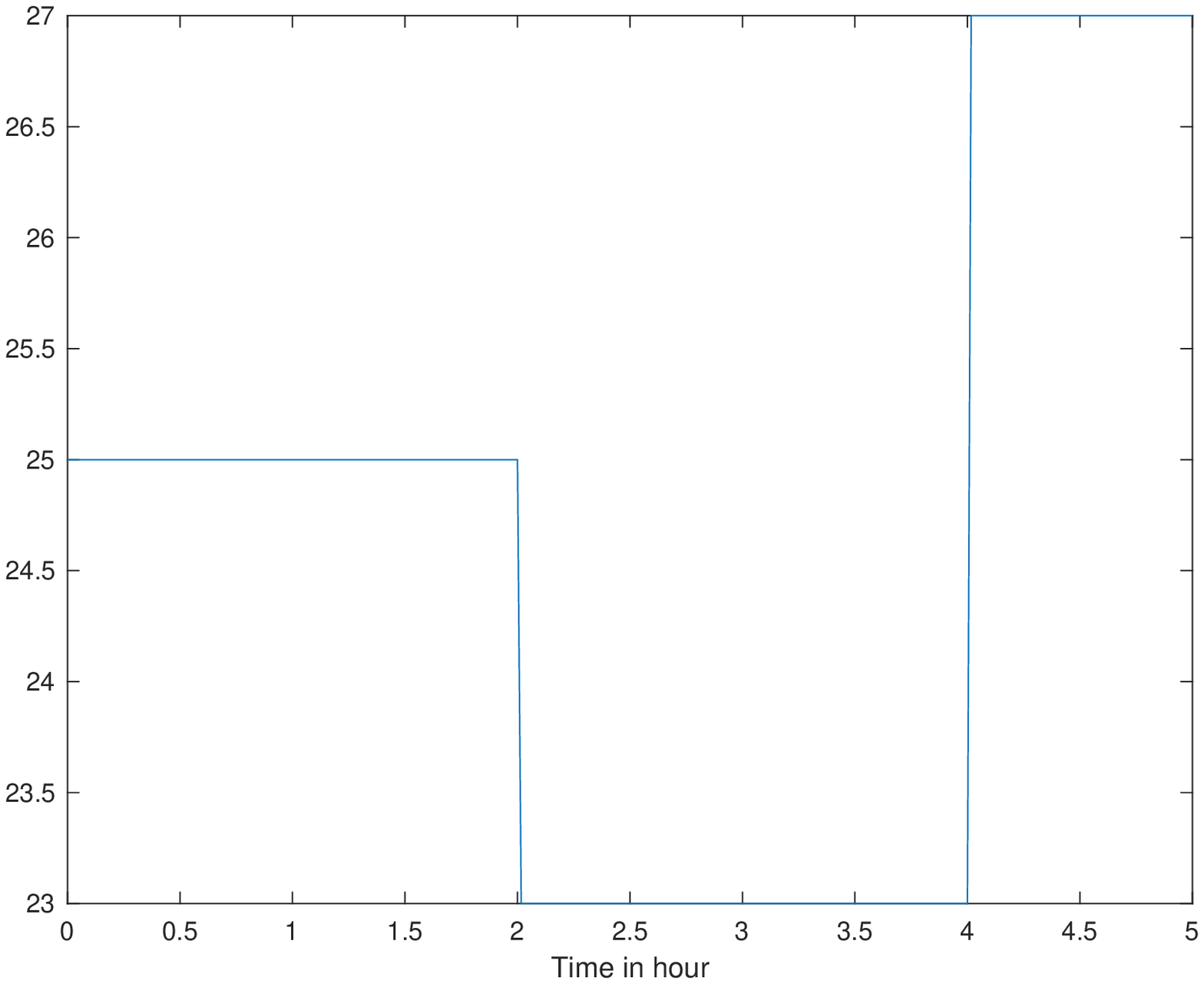,width=0.30\textwidth}}%\hspace{.5cm}
\\
\subfigure[\footnotesize Control variable $T_{Air,in}$ (degree Celsius) ]
{\epsfig{figure=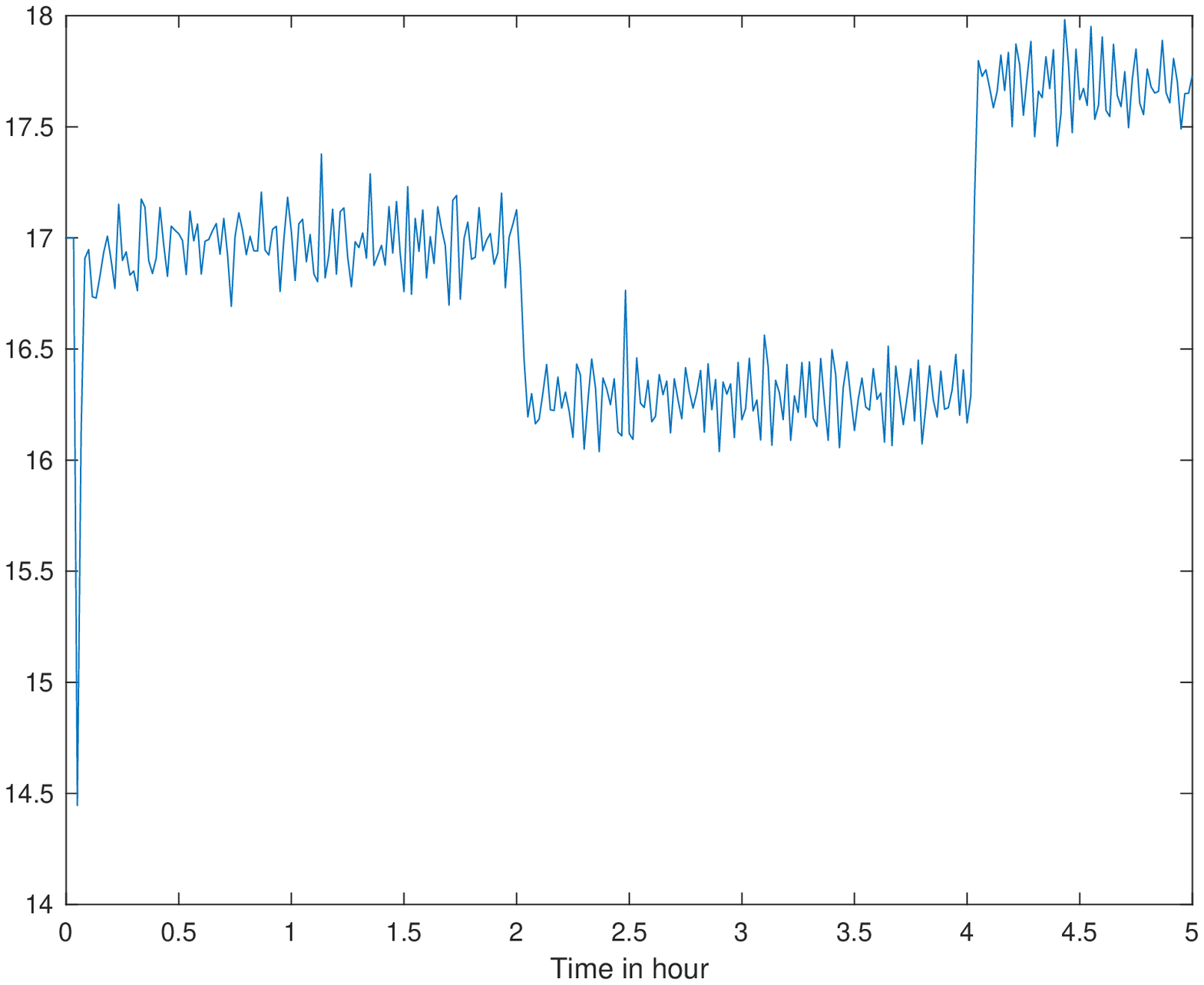,width=0.30\textwidth}}%\hspace{.5cm}
\subfigure[\footnotesize Output variable $T_{IT}$ (degree Celsius) ]
{\epsfig{figure=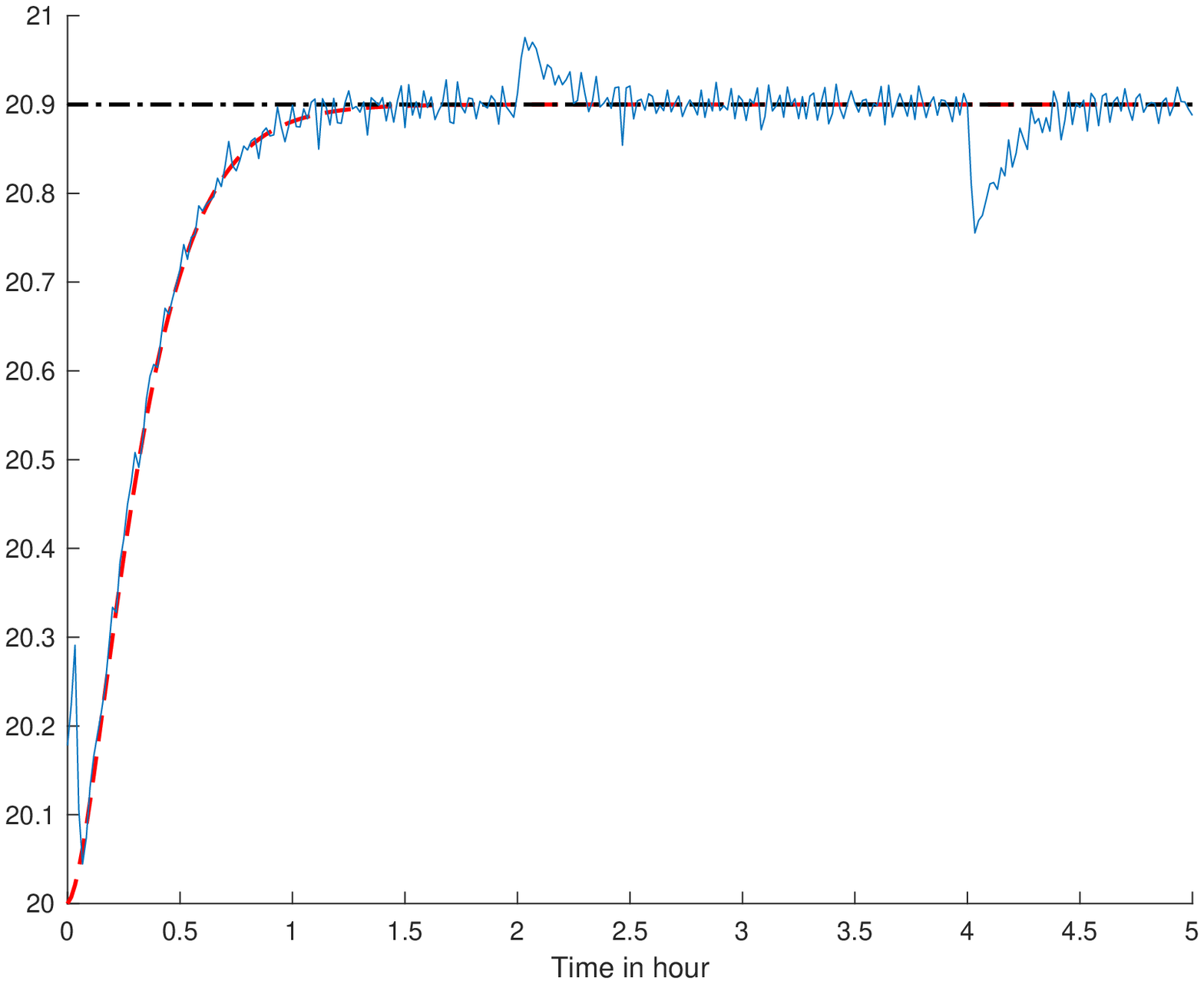,width=0.30\textwidth}}%\hspace{.5cm}
\caption{Sudden temperature change}\label{S1}
\end{figure*}

Figure \ref{S1}-(b) exhibits a sudden temperature of the data center temperature $T_{\rm out}$. Here again Figure \ref{S1}-(d) depicts an excellent tracking.

\subsubsection{Another reference trajectory}
\begin{figure*}[!ht]
\centering%
\subfigure[\footnotesize Power consumption: CPU load $P_{IT}$  (kW) ]
{\epsfig{figure=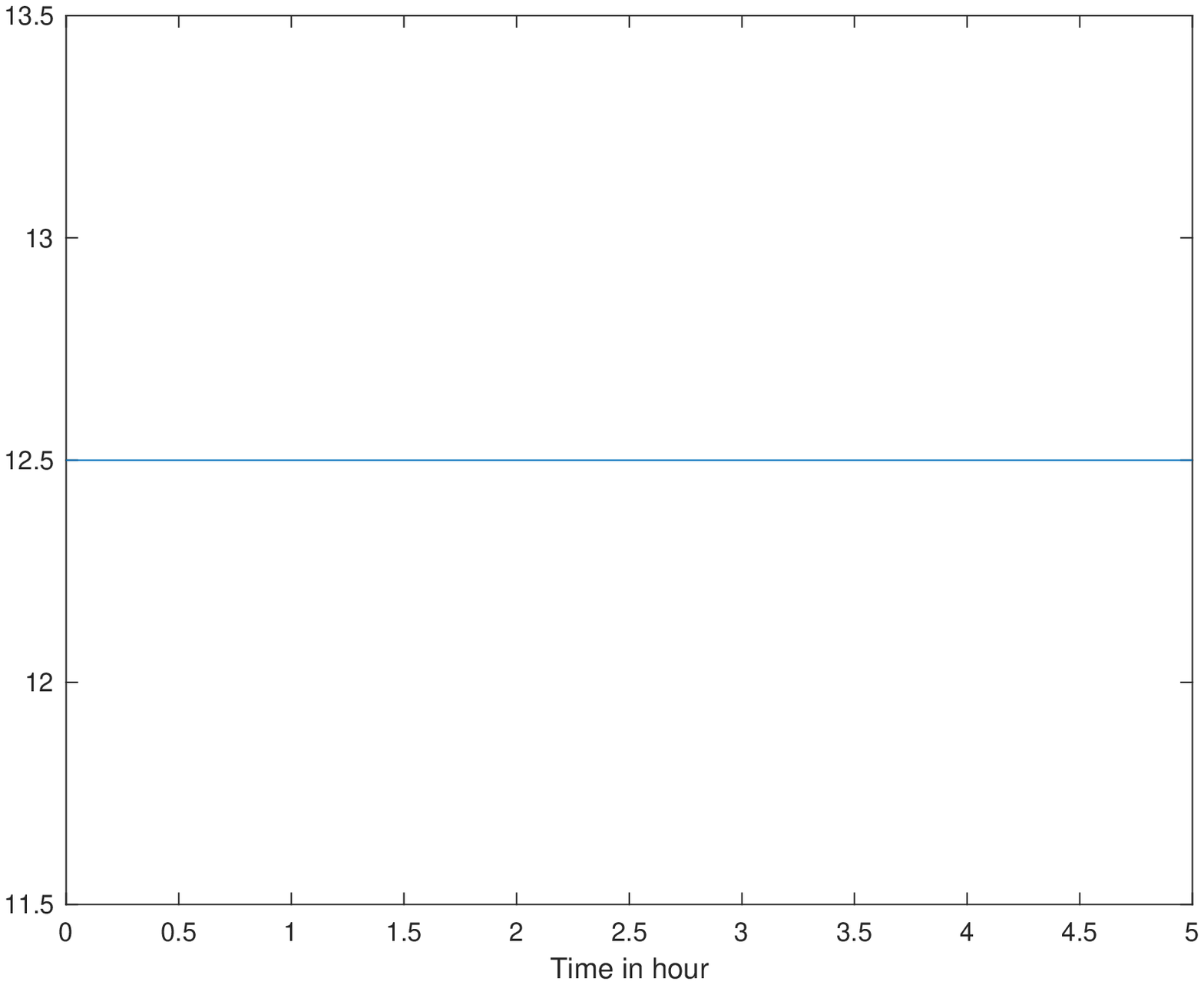,width=0.30\textwidth}}%\hspace{.5cm}
%
%\\
\subfigure[\footnotesize Data center temperature $T_{out}$ (degree Celsius) ]
{\epsfig{figure=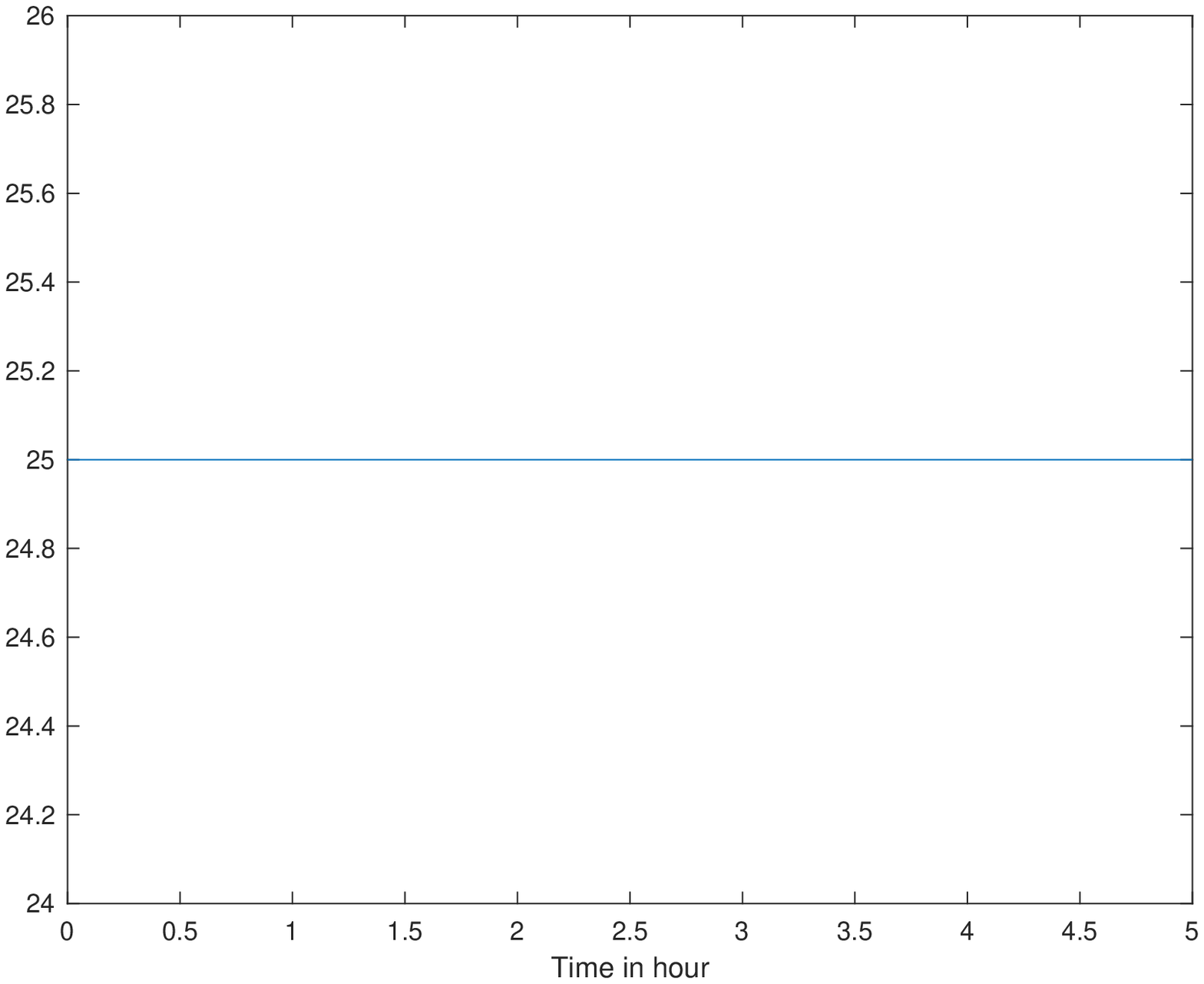,width=0.30\textwidth}}%\hspace{.5cm}
\\
\subfigure[\footnotesize Control variable $T_{Air,in}$ (degree Celsius)
 ]
{\epsfig{figure=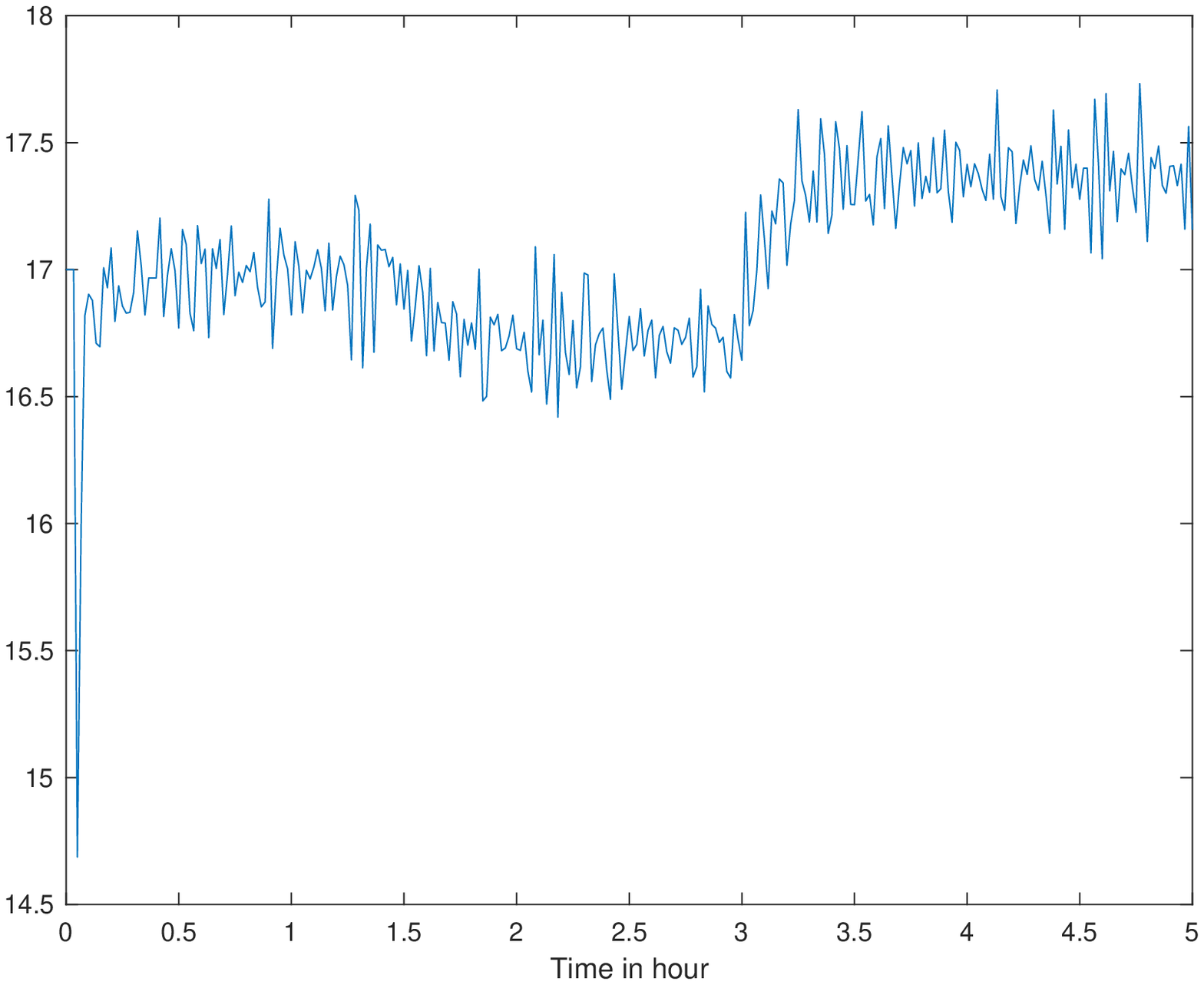,width=0.30\textwidth}}%\hspace{.5cm}
\subfigure[\footnotesize Output variable $T_{IT}$ (degree Celsius)]
{\epsfig{figure=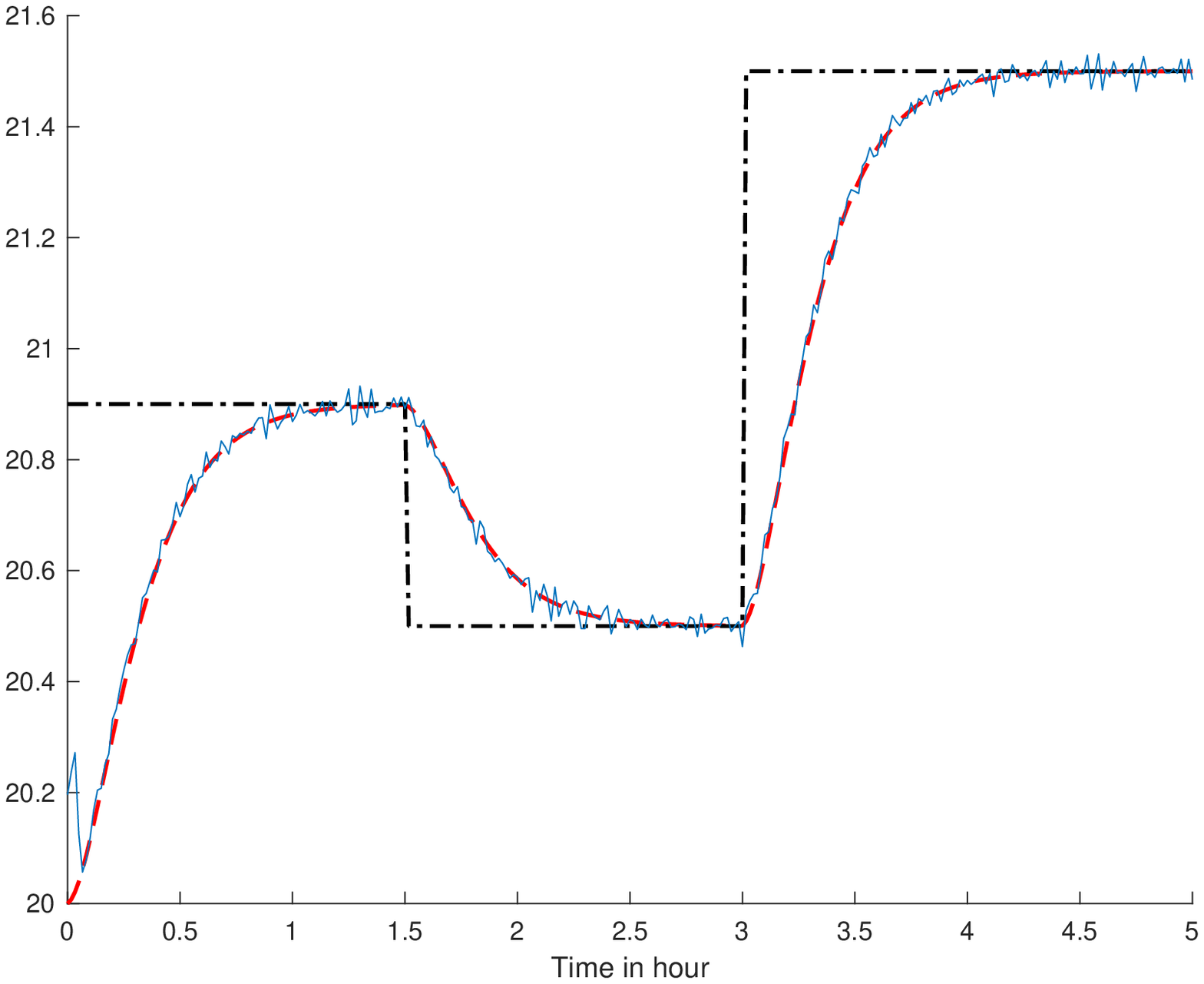,width=0.30\textwidth}}%\hspace{.5cm}
\caption{Another reference trajectory}\label{S4}
\end{figure*}

Some situations may necessitate, contrarily to Section \ref{basic}, to replace the setpoint, \textit{i.e.}, a constant reference trajectory, by a more general one. As demonstrated by Figure \ref{S4}, the tracking remains exceptional.

%Ainsi, on cherche alors à maitriser la temp\'{e}rature $T_{IT}$ (aussi not\'{e}e $y$) proche de $20.9°c$\footnote{Temp\'{e}rature de fonctionnement recommand\'{e}e d'après l'article.} en agissant sur $T_{Air,in}$ (aussi not\'{e}e $u$) . La principale perturbation est %la variation de charge des processeurs $P_{IT}$. Nous \'{e}valuons aussi la robustesse de notre proposition à une variation de temp\'{e}rature $T_{out}$.\\
%Les simulations sont men\'{e}es avec une p\'{e}riode d'\'{e}chantillonnage de $1min$ et $\alpha=10$, $K_p=-1$ avec le modèle ultra-local $$\dot y=F+\alpha u$$ la commande associ\'{e}e $$u=\frac{1}{\alpha}\left(-\hat F+K_p e +\dot y^\ast \right)$$ où $e=y-%y^\ast$ et $y^\ast$ la r\'{e}f\'{e}rence.

\subsection{Sudden model change}
\begin{figure*}[!ht]
\centering%
%\subfigure[\footnotesize Charge processeur $P_{IT}$  en {$[ kW ]$}]
%{\epsfig{figure=C5.eps,width=0.30\textwidth}}%\hspace{.5cm}
%%
%%\\
%\subfigure[\footnotesize Temperature du Data Center $T_{out}$  en {$[ °c ]$} ]
%{\epsfig{figure=Tout5.eps,width=0.30\textwidth}}%\hspace{.5cm}
%%
%\\
\subfigure[\footnotesize Control variable $T_{Air,in}$ (degree Celsius) ]
{\epsfig{figure=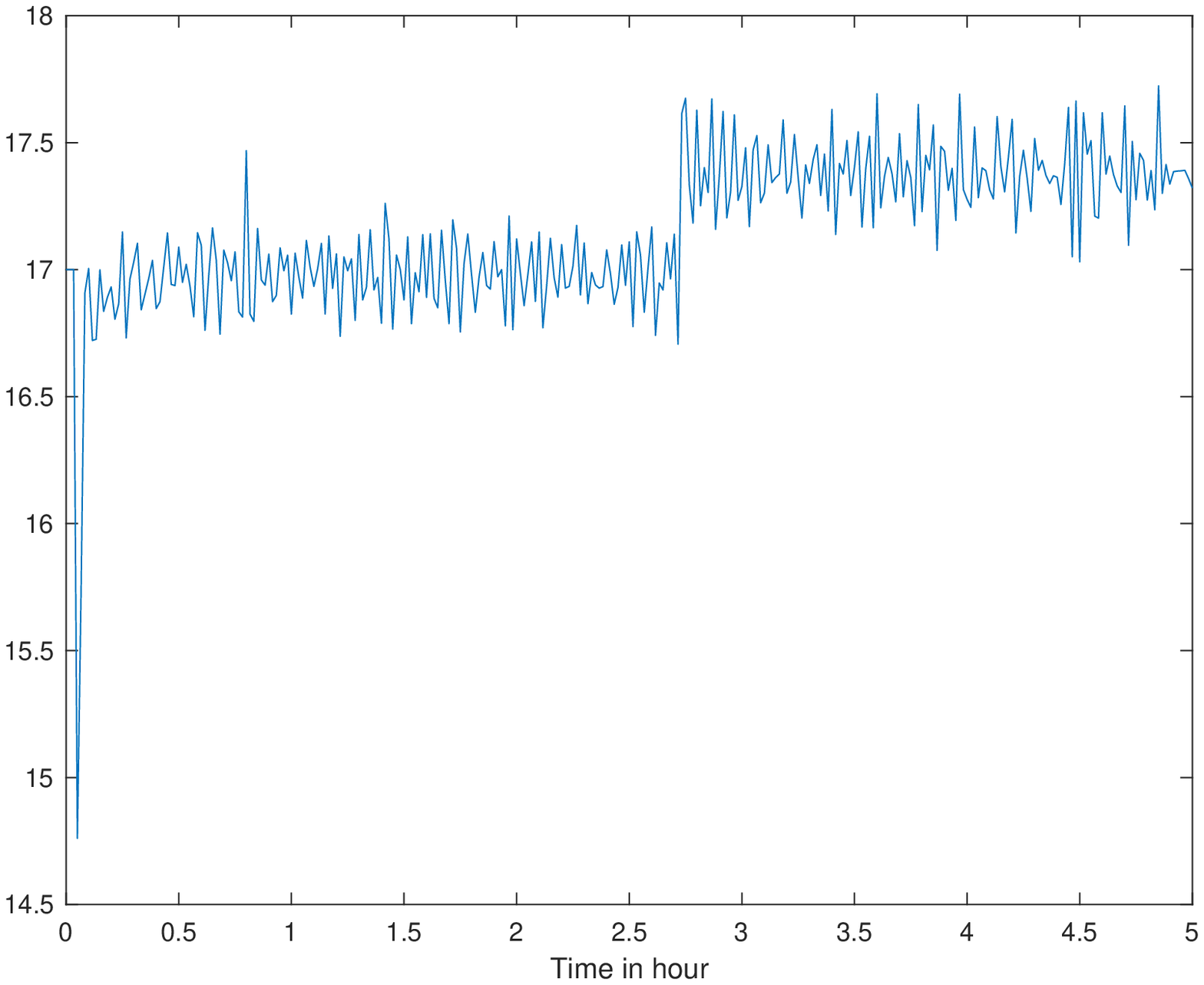,width=0.30\textwidth}}%\hspace{.5cm}
\subfigure[\footnotesize Output variable $T_{IT}$ (degree Celsius)]
{\epsfig{figure=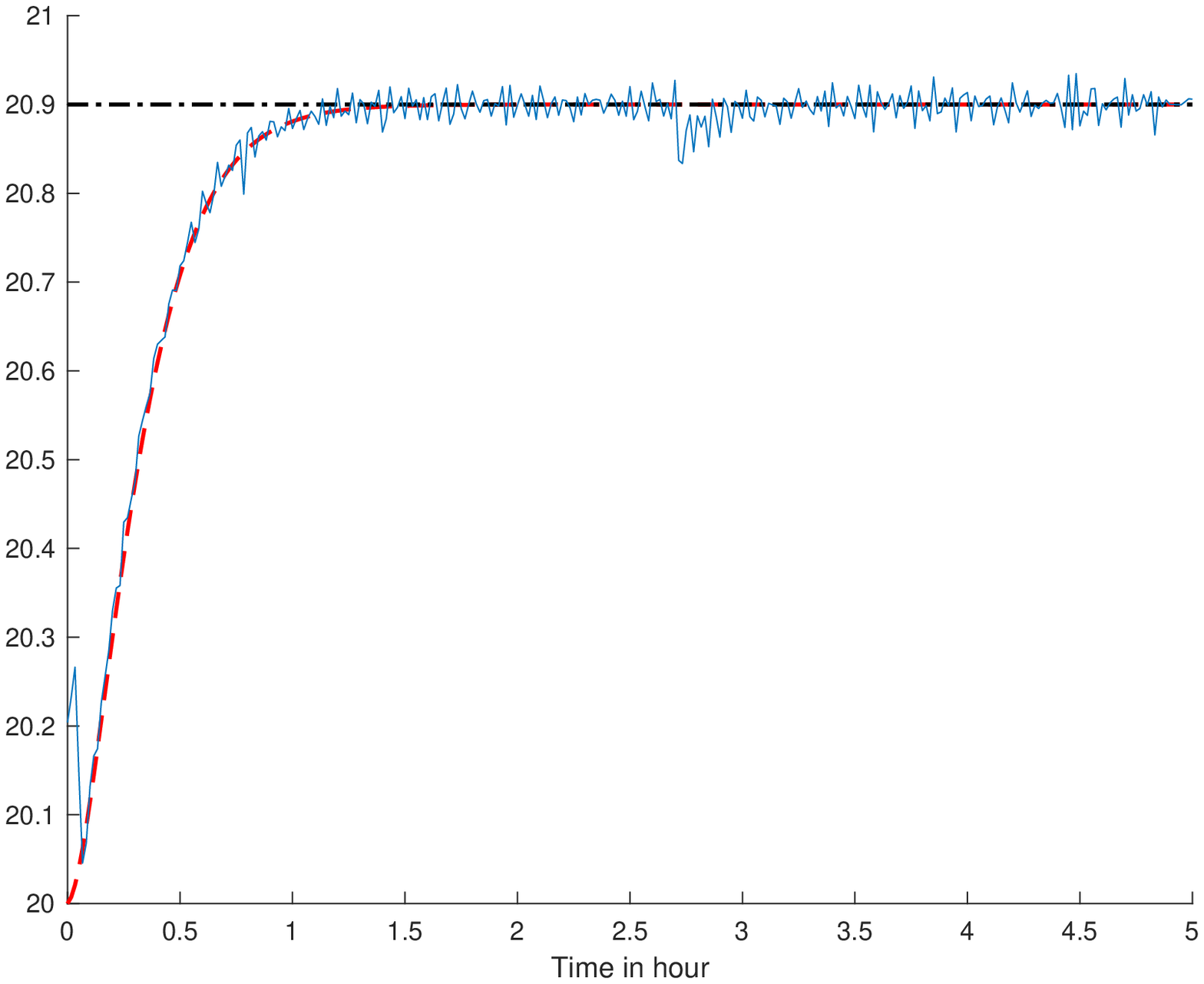,width=0.30\textwidth}}%\hspace{.5cm}
\caption{Parameters change: $\times 1.5$}\label{S5}
\end{figure*}

\begin{figure*}[!ht]
\centering%
%\subfigure[\footnotesize Charge processeur $P_{IT}$  en {$[ kW ]$}]
%{\epsfig{figure=C6.eps,width=0.30\textwidth}}%\hspace{.5cm}
%%
%%\\
%\subfigure[\footnotesize Temperature du Data Center $T_{out}$  en {$[ °c ]$} ]
%{\epsfig{figure=Tout6.eps,width=0.30\textwidth}}%\hspace{.5cm}
%%
%\\
\subfigure[\footnotesize Control variable $T_{Air,in}$ (degree Celsius) ]
{\epsfig{figure=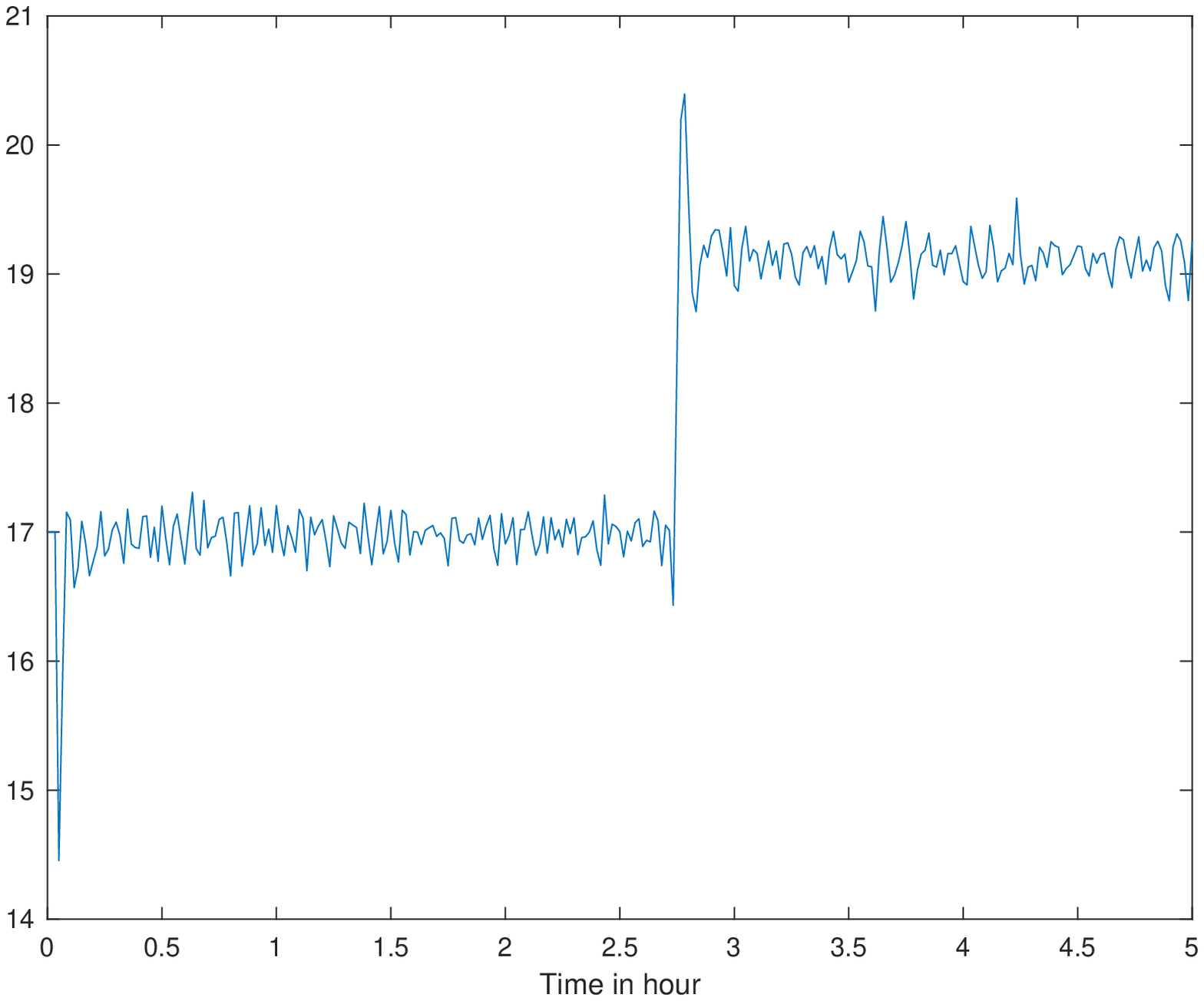,width=0.30\textwidth}}%\hspace{.5cm}
\subfigure[\footnotesize Output variable $T_{IT}$ (degree Celsius)]
{\epsfig{figure=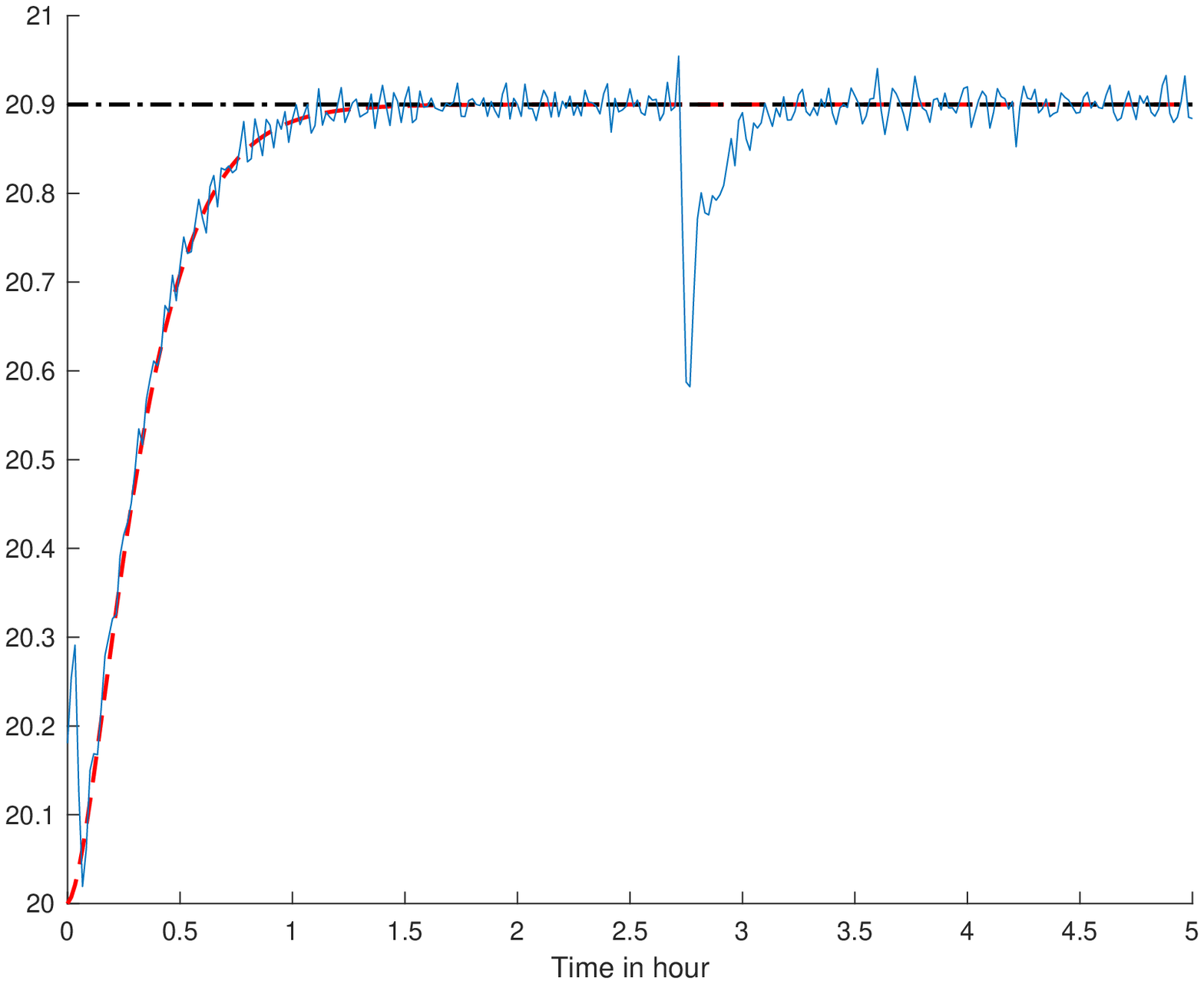,width=0.30\textwidth}}%\hspace{.5cm}
\caption{Parameter change: $\times 0.5$}\label{S6}
\end{figure*}

Represent a sudden model change at time $t=2.7$ h by multiplying $\alpha_{21}$, $\alpha_{31}$, $\alpha_{51}$ in Equations \eqref{syst} by $0.5$ and $1.5$.\footnote{In accordance with \cite{cupelli}, those variations may be justified by the change of a coefficient called $\kappa$.} If those changes would occur at time $t =0$, they should be interpreted as a model mismatch. The variables $P_{\rm IT}$ and $T_{out}$ remain unaltered and constant. Although the model-free control synthesis of 
Section \ref{basic} remains unchanged, Figures \ref{S5} et \ref{S6} display excellent performances.

%Cette variation se justifie par le fait qu'un paramètre nomm\'{e} $\kappa$ dans In accordance with \cite{cupelli}, this can be explained by a change of a parameter  intervient dans ces trois coefficients et est utilis\'{e} pour ajuster le modèle. Une variation %de ce paramètre peut donc être ais\'{e}ment interpr\'{e}t\'{e}e comme une erreur de mod\'{e}lisation ou une variation de modèle.
%Pour ces deux simulations, la charge processeur et la temp\'{e}rature du data center sont identiques à celles pr\'{e}sent\'{e}es en \ref{S4}-(a) et \ref{S4}-(b).

%%%%%%%%%%%%%%%%%

\section{Conclusion} \label{con}
The \emph{power usage effectiveness}, or \emph{PUE}, of data centers, although heavily criticized \cite{pue}, seems to be the only measure for checking the energy saving quality today. It would however be meaningless 
to try applying this indicator here, in the context of such a paper. Section \ref{simu}, which demonstrates that the tracking works well in rather stringent conditions, may convince the reader that our approach should be nevertheless 
quite efficient with respect to energy saving. The most important for future developments is of course the application of our method to real data. A positive outcome would lead to a critical simplification of the HVAC control management of data centers:
\begin{itemize}
\item irrelevance of complex and time-consuming mathematical modeling, which is inherently uncertain,
\item forthright tuning.
\end{itemize}
Promising experiments with a greenhouse \cite{toulon} and a building \cite{micha} comfort this hope. It would confirm thanks also to \cite{iste} that model-free control should become important in computer science (compare with \cite{astrom,hellerstein,janert}).

\end{document}